\documentclass{article}
\usepackage{spconf,amsmath,graphicx}
\usepackage{booktabs}
\usepackage{enumitem}
\setlist{nosep, leftmargin=14pt}

\usepackage{mwe} % to get dummy images
\usepackage{hyperref}
\hypersetup{colorlinks,linkcolor={blue},citecolor={blue},urlcolor={blue}}
\title{State-of-the-Art Stroke Lesion Segmentation at 1/1000th of Parameters}
\name{
\begin{tabular}{@{}c@{}}
Alex Fedorov$^{\clubsuit\dagger}$ \qquad
Yutong Bu$^{\diamondsuit\dagger}$ \qquad
Xiao Hu$^{\clubsuit\dagger}$ \qquad
Chris Rorden$^{\spadesuit}$ \qquad
Sergey Plis$^{\heartsuit}$ \qquad
\end{tabular}
}
\address{
$^{\clubsuit}$ Center for Data Science, Nell Hodgson Woodruff School of Nursing, \\
$^{\diamondsuit}$ Department of Mathematics, $^{\dagger}$ Emory University, Atlanta, GA, USA\\
$^{\spadesuit}$ Department of Psychology, University of South Carolina, Columbia, SC, USA \\
$^{\heartsuit}$ Department of Computer Science, Georgia State University, Atlanta, GA, USA
}

\begin{document}

\maketitle

\begin{abstract}
Efficient and accurate whole-brain lesion segmentation remains a
challenge in medical image analysis. In this work, we revisit MeshNet,
a parameter-efficient segmentation model, and introduce a novel
multi-scale dilation pattern with an encoder-decoder structure. This innovation enables capturing broad contextual information and fine-grained details without traditional downsampling, upsampling, or skip-connections. Unlike previous approaches processing subvolumes or slices, we operate directly on whole-brain \(256^3\) MRI volumes. Evaluations on the Aphasia Recovery Cohort (ARC) dataset demonstrate that MeshNet achieves superior or comparable DICE scores to state-of-the-art architectures such as MedNeXt and U-MAMBA at 1/1000th of parameters. Our results validate MeshNet's strong balance of efficiency and performance, making it particularly suitable for resource-limited environments such as web-based applications and opening new possibilities for the widespread deployment of advanced medical image analysis tools.
\end{abstract}

\begin{keywords}
Chronic Stroke, Lesion Segmentation, Whole-Brain Segmentation, Magnetic Resonance Imaging
\end{keywords}

\section{Introduction}

Accurate segmentation of brain lesions in MRI can aid in the diagnosis, treatment planning, and monitoring of neurological conditions, including chronic stroke. While manual lesion delineation is often subjective and time-consuming~\cite{de2015fast}, automatic segmentation offers a more rapid and objective solution. The state-of-the-art architectures such as MedNeXt~\cite{roy2023mednext} and U-MAMBA~\cite{ma2024u}, based on the U-Net~\cite{ronneberger2015u}, deliver high accuracy. Yet these advances come at the cost of increased complexity or computational demands. Such large models are challenging to deploy in resource-limited or edge-based environments and may necessitate sending sensitive patient data to remote servers.

Several 3D segmentation models improved U-Net designs by integrating residuals, volumetric convolutions, and transformers to capture multi-scale features, as seen in architectures such as V-Net~\cite{milletari2016v}, SegResNet~\cite{myronenko20193d}, Swin-UNETR~\cite{hatamizadeh2021swin}, and MedNeXt~\cite{roy2023mednext}. Emerging approaches like U-MAMBA~\cite{ma2024u} and U-KAN~\cite{li2024u} combine U-Net with novel MAMBA and KAN blocks. These models are generally similar to U-Net and have been shown to improve their performance with novel, sophisticated blocks. However, they still require substantial computational resources and memory, making them challenging to deploy in resource-limited environments.

    To address these limitations, we revisit MeshNet~\cite{fedorov2017end}, a fully convolutional architecture with a low parameter count, and enhance it with a novel dilation pattern to capture both broad contextual and fine-grained details. We perform directly whole-brain $256^3$ MRI volume segmentation, without relying on 3D subvolume or 2D slice sampling, and benchmark our model on the Aphasia Recovery Cohort (ARC) dataset~\cite{gibson2024aphasia}. Our experiments demonstrate segmentation performance that rivals or surpasses current state-of-the-art models with 1,000 times more parameters. Furthermore, we demonstrate its balance of parameter complexity and segmentation performance.

\begin{table*}[t]
\centering
\caption{Model performance metrics with parameter counts. AVD denotes
  Average Volume Difference and MCC denotes Matthews's Correlation
  Coefficient. (*) indicate models statistically significantly different
  from MeshNet-26 ($p < 0.05$, Holm-corrected Wilcoxon test).}
\label{tab:Raw_metrics_params}
{\small
\begin{tabular}{|l|c|c|c|c|c|c|c|}
\toprule
Model & Parameters & DICE (↑) & $p<0.05$ & AVD (↓) & $p<0.05$ & MCC (↑) & $p<0.05$ \\
\midrule
MeshNet-26 & 147 474 & 0.876 (0.016) & N/A & 0.245 (0.036) & N/A & 0.760 (0.030) & N/A \\
MeshNet-16 & 56 194 & 0.873 (0.007) &  & 0.249 (0.033) &  & 0.757 (0.013) &  \\
U-MAMBA-BOT & 7 351 400 & 0.870 (0.012) &  & 0.266 (0.053) &  & 0.750 (0.023) &  \\
MedNeXt-M & 17 548 963 & 0.868 (0.017) &  & 0.288 (0.094) &  & 0.745 (0.033) &  \\\midrule
SegResNet & 1 176 186 & 0.867 (0.005) &  & 0.268 (0.039) & * & 0.743 (0.011) &  \\
U-MAMBA-ENC & 7 514 280 & 0.865 (0.010) &  & 0.309 (0.080) & * & 0.740 (0.022) &  \\
MedNeXt-S & 5 201 315 & 0.861 (0.012) &  & 0.257 (0.054) &  & 0.729 (0.021) &  \\
Swin-UNETR & 18 346 844 & 0.859 (0.025) &  & 0.425 (0.283) & * & 0.730 (0.048) &  \\
\midrule
MedNeXt-B & 10 526 307 & 0.855 (0.016) & * & 0.260 (0.039) & * & 0.718 (0.030) & * \\
U-KAN & 44 070 082 & 0.851 (0.013) & * & 0.321 (0.027) & * & 0.717 (0.025) & * \\
MeshNet-5 & 5 682 & 0.848 (0.023) & * & 0.280 (0.060) & * & 0.708 (0.042) & * \\
UNETR & 95 763 682 & 0.847 (0.014) & * & 0.397 (0.029) & * & 0.700 (0.027) & * \\
Residual U-Net & 1 979 610 & 0.836 (0.005) & * & 0.562 (0.495) & * & 0.685 (0.010) & * \\
V-Net & 45 597 898 & 0.798 (0.018) & * & 0.387 (0.047) & * & 0.614 (0.033) & * \\
\bottomrule
\end{tabular}
    \vspace{-10pt}
}
\end{table*}

\section{Methods}
\paragraph*{MeshNet Revisited}

The original MeshNet architecture~\cite{fedorov2017end} is a fully convolutional segmentation architecture, leveraging dilated kernels to expand the receptive field. It employed a steadily increasing dilation pattern of \(1 \to 1\!\to\!2\!\to\!4\!\to\!8\!\to\!16\!\to\!1\!\to\!1\) across 8 convolutional layers. This approach allowed the model to capture larger spatial contexts without significantly increasing the number of parameters. However, this pattern was limited in its ability to recover finer spatial details, as it did not decrease the dilation rate in later layers, restricting the model’s flexibility in refining segmentation boundaries. Consequently, while the original MeshNet provided an efficient solution for context integration, it faced challenges balancing global context capture and spatial detail preservation.

Our revisited MeshNet architecture refines the original design by adopting an expanded 10-layer structure with an adaptive dilation pattern. This pattern starts with progressively increasing dilation values—\(1\!\to\!2\!\to\!4\!\to\!8\!\to\!16\)—to expand the receptive field and capture broad contextual information. In the latter layers, the dilation values symmetrically decrease—\(16\!\to\!8\!\to\!4\!\to\!2\!\to\!1\)—allowing for a gradual transition back to smaller receptive fields, which enhances spatial detail and context integration. This design mimics the information flow seen in U-Net's encoder-decoder structure without relying on traditional downsampling, upsampling, or skip connections.

Unlike typical U-Net-based architectures, MeshNet maintains multi-scale context solely through controlled dilation and padding, eliminating the need for feature concatenation between encoder and decoder layers. Each layer’s receptive field is carefully adjusted to ensure spatial consistency across scales. This approach significantly reduces memory requirements by avoiding the storage of high-dimensional feature maps from earlier layers, making the model more memory-efficient during inference.

We explored three variants of MeshNet-X, where $X$ represents the number of channels: $X = 5$ (5,682 parameters), $X = 16$ (56,194), and $X = 26$ (147,474). With its low parameter count, reducing the storage requirements for weights, and its efficient layer-by-layer inference, the MeshNet design minimizes computational overhead, making it highly suitable for high-resolution segmentation on resource-constrained hardware and for browser-based MRI segmentation
\href{brainchop.org}{brainchop.org}~\cite{masoud2023}, where low memory usage and computational efficiency are critical.

\paragraph*{Dataset} The Aphasia Recovery Cohort (ARC)~\cite{gibson2024aphasia} is an open-source neuroimaging dataset comprising T2-weighted MRI scans from 230 unique individuals with chronic stroke. To prepare the dataset for segmentation analysis, we implemented several preprocessing steps. Using \texttt{mri\_convert}~\cite{dale1999cortical}  tool, we resampled all images to a uniform \(1\)mm isotropic resolution with dimensions of \(256 \times 256 \times 256\) voxels and applied a 0-1 rescaling to normalize intensity values across images. We utilized Sampling Perfection with Application optimized Contrast using different flip angle Evolution (SPACE) sequences with x2 in plane acceleration (115 scans) and without acceleration (109 scans), while excluding the turbo-spin echo T2-weighted sequences (5 scans) to maintain homogeneity in imaging protocols. Unlike standard approaches that use 3D subvolume sampling or 2D slices, we use whole-brain \(256^3\) cubes for both training and inference, ensuring full-brain coverage in each pass.

We implemented a nested cross-validation approach with three outer folds containing a train and hold-out test set. We further divided each outer fold's training data into three inner folds, creating a robust nested structure for training and validation sets. We calculated lesion volumes as a number of voxels to stratify dataset splits across different lesion sizes and acquisition types to ensure each fold preserved a representative distribution. We categorized the images based on quintiles of the lesion size distribution. We used the 0th, 25th, 50th, 75th, and 100th percentiles as cutoff points. The distribution ranges are: $(203, 33619]$, $(33619, 67891]$, $(67891, 128314]$, and $(128314, 363885]$.

\paragraph*{Training} For model training of the baselines, we used the AdamW optimizer with a learning rate of $0.001$, weight decay of $3\text{e-}5$, and epsilon of $1\text{e-}4$ with a batch size of 1. We implemented a OneCycleLR learning rate scheduler. The scheduler starts at 1/100th of the max learning rate, quickly increases to the maximum, and gradually decreases. The maximum learning rate was set to $0.001$, with the increase phase occurring in the first $1\%$ of training time. We trained the model for $50$ epochs. The training has been performed using half-precision. The objective is cross-entropy loss with label smoothing of 0.01 and class weighting of 0.5 for the background and 1.0 for lesions.

For the training of MeshNet, we followed the same procedure. However, the hyperparameters are chosen based on the hyperparameter search, and we trained the model with $10$
restarts. To optimize hyperparameters of MeshNet, we conducted a hyperparameter search using Oríon~\cite{xavier_bouthillier_2023_0_2_7}, an asynchronous framework for black-box function optimization. We employed the Asynchronous Successive Halving Algorithm (ASHA). It allows us to initially evaluate configurations with fewer epochs and promote promising ones to higher fidelity levels. Our hyperparameter search space included: number of channels $\text{uniform~(5, 21)}$, learning rate as $\text{log uniform~(1e-4, 4e-2)}$, weight decay as $\text{loguniform~(1e-4, 4e-2)}$, background weight as $\text{uniform~(0, 1)}$, percentage of warmup as $\text{[0.02, 0.1, 0.2]}$, and number of epochs as $\text{fidelity~(15, 50)}$. Hyperparameter optimization was conducted on the inner folds of the first outer fold. The optimized hyperparameters were then applied to train models on all outer folds.

The learning framework was implemented using PyTorch with the Hydra configuration system. Each experiment was conducted on a single NVIDIA A100 GPU with 80 GB memory. We used layer checkpointing for models that could not fit into GPU memory.

\begin{figure}[t]
    \centering
    \includegraphics[width=0.95\linewidth]{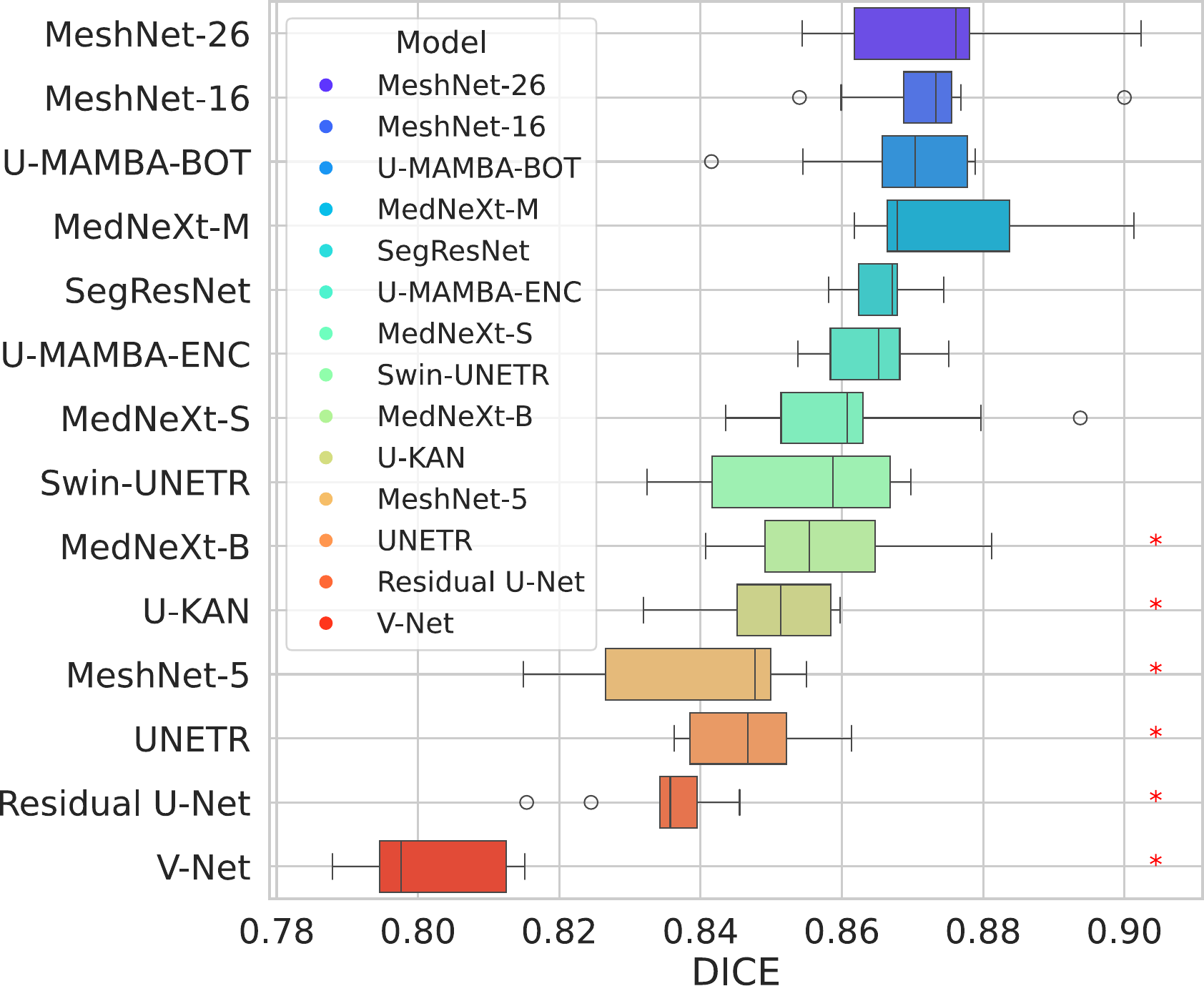}
    \vspace{-10pt}
    \caption{Comparison of DICE scores across models for whole-brain lesion segmentation. Models are ranked by median DICE score, with MeshNet-26 achieving the highest performance. Red asterisks (*) indicate models with statistically significant differences ($p < 0.05$, Holm-corrected Wilcoxon test) compared to MeshNet-26.}
    \label{fig:Raw_DICE_loss}
    \vspace{-20pt}
\end{figure}

\section{Results}

\paragraph*{MeshNet rivals larger models on the ARC dataset}

Table~\ref{tab:Raw_metrics_params} presents the performance metrics of various models along with their parameter counts. The MeshNet variants, particularly MeshNet-26 and MeshNet-16, demonstrate superior performance among all models tested. MeshNet-26 achieves the highest DICE coefficient of $0.876~(0.016)$ with $147,474$ parameters. MeshNet-16, with a smaller parameter count of $56,194$, also performs exceptionally well, attaining a DICE of $0.873~(0.007)$. Both models exhibit strong MCC values—$0.760~(0.030)$ for MeshNet-26 and $0.757~(0.013)$ for MeshNet-16—indicating reliable segmentation performance with relatively low computational complexity.

Other models, such as U-MAMBA-BOT and MedNeXt-M, also show competitive results but with significantly higher parameter counts. With over 7 million parameters, U-MAMBA-BOT achieves a DICE coefficient of $0.870~(0.012)$ and an MCC of $0.750~(0.023)$. MedNeXt-M, having more than 17 million parameters, attains a DICE of $0.868~(0.017)$ and an MCC of $0.745~(0.033)$. While these models perform well, their larger sizes make them less efficient than the MeshNet variants. Models like SegResNet and MedNeXt-S balance performance and parameter count but do not surpass the MeshNet models in key metrics.

Lower-performing models include MedNeXt-B, U-KAN, MeshNet-5, UNETR, Residual U-Net, and V-Net. MeshNet-5, the smallest variant with only 5,682 parameters, achieves a DICE of $0.848~(0.023)$ and an MCC of $0.708~(0.042)$, which is significant given its minimal size but lower than its larger counterparts. UNETR and V-Net, despite having substantial parameter counts exceeding 45 million, deliver lower DICE coefficients of $0.847~(0.014)$ and $0.798~(0.018)$, respectively. These results suggest that a higher number of parameters alone does not necessarily correlate with better performance and highlights the parameter efficiency of the MeshNet architecture.

\begin{figure}[t]
    \centering
    \includegraphics[width=0.95\linewidth]{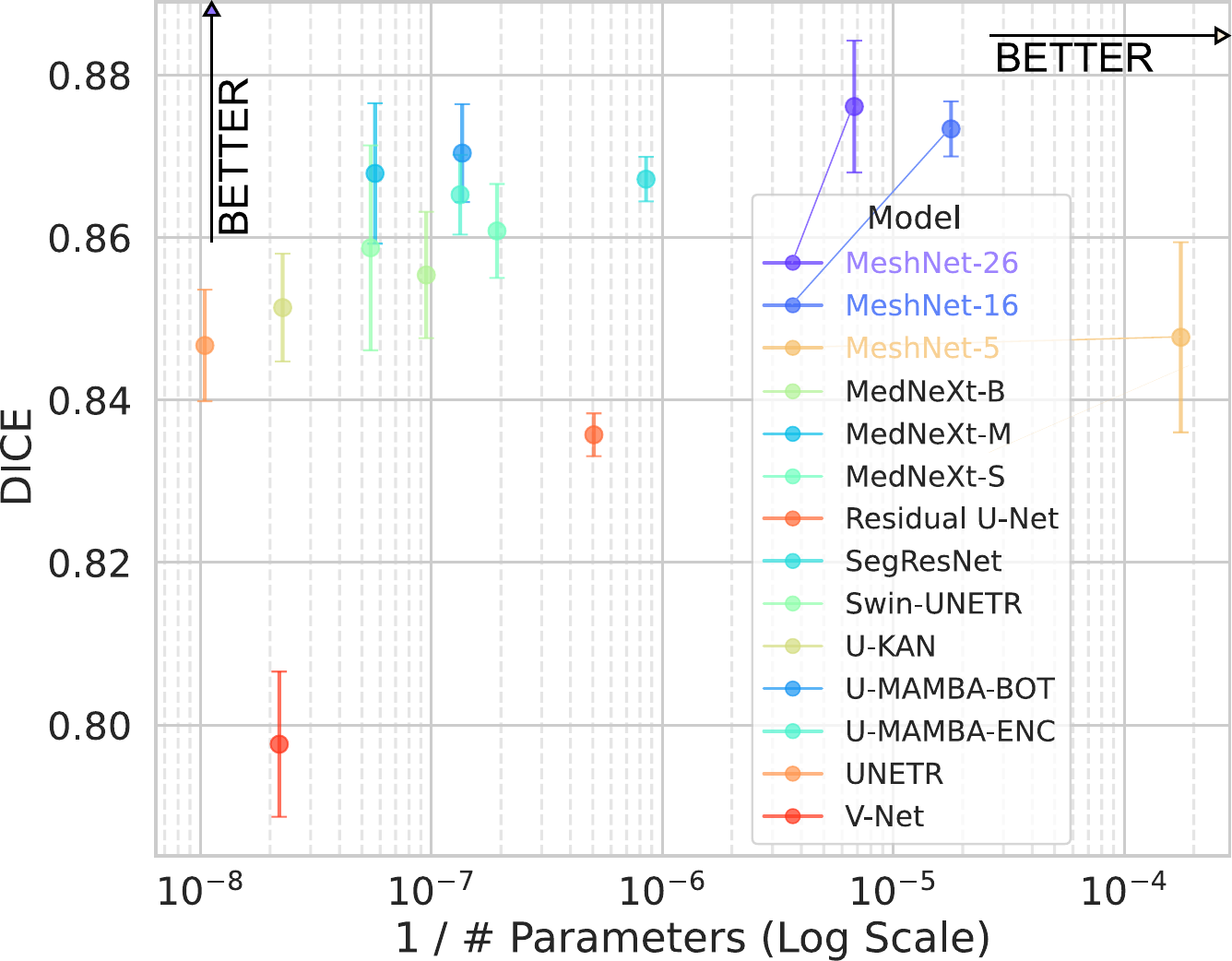}
    \vspace{-10pt}
    \caption{Relationship between model complexity (represented as the
      the inverse of parameter count on a log scale) and median DICE score
      with interquartile range (IQR) error bars. Models with fewer
      parameters appear toward the right, while models with higher
      parameter counts are positioned on the left. MeshNet models are
      on the Pareto frontier indicating the balance between parameter
      efficiency and accuracy. The highest number of channels (MeshNet-26) produces
      the best DICE.
    }
    \vspace{-10pt}
    \label{fig:Raw_params}
\end{figure}

\paragraph*{MeshNet balances high performance and low parameter count}
Figure~\ref{fig:Raw_params} presents a scatter plot illustrating the
trade-off between model complexity and segmentation performance across
different architectures. The x-axis displays the inverse of the
parameter count on a log scale, positioning models with larger parameter savings toward the right. The y-axis represents the median DICE score, with interquartile range (IQR) error bars indicating variability in segmentation accuracy. Models achieving higher DICE scores are located higher on the plot. Notably, MeshNet configurations lie on the Pareto frontier, demonstrating an optimal balance between parameter efficiency and segmentation accuracy.

MeshNet-26 demonstrates remarkable parameter efficiency, achieving a DICE score of 0.876 using only $147,474$ parameters, making it about $50\times$ and $120\times$ more efficient than next best performant models U-MAMBA-BOT (0.870 DICE) and MedNeXt-M (0.868 DICE), respectively. Similarly, MeshNet-16 achieves a DICE of 0.873 with just $56,194$ parameters, making it $130\times$ and $310\times$ more efficient than U-MAMBA-BOT and MedNeXt-M.

The ultra-compact MeshNet-5 model, with 5,682 parameters, achieves a DICE of 0.848, slightly below the 0.85 threshold needed for reliable segmentation~\cite{liew2022large}. It is $1300\times$ more efficient than U-MAMBA-BOT and $3089\times$ more efficient than MedNeXt-M. Other architectures with comparable performance, such as UNETR (DICE 0.847), U-KAN (DICE 0.851), and MedNeXt-B (DICE 0.855), require significantly more parameters: approximately $784\times$, $1704\times$, and $187\times$, respectively. The results highlight MeshNet-5’s substantial parameter efficiency relative to these architectures. This also suggests a potential for an intermediate model between MeshNet-5 and MeshNet-16 to optimize parameter efficiency further while maintaining robust accuracy.

\begin{figure}[t]
    \centering
    \includegraphics[width=0.95\linewidth]{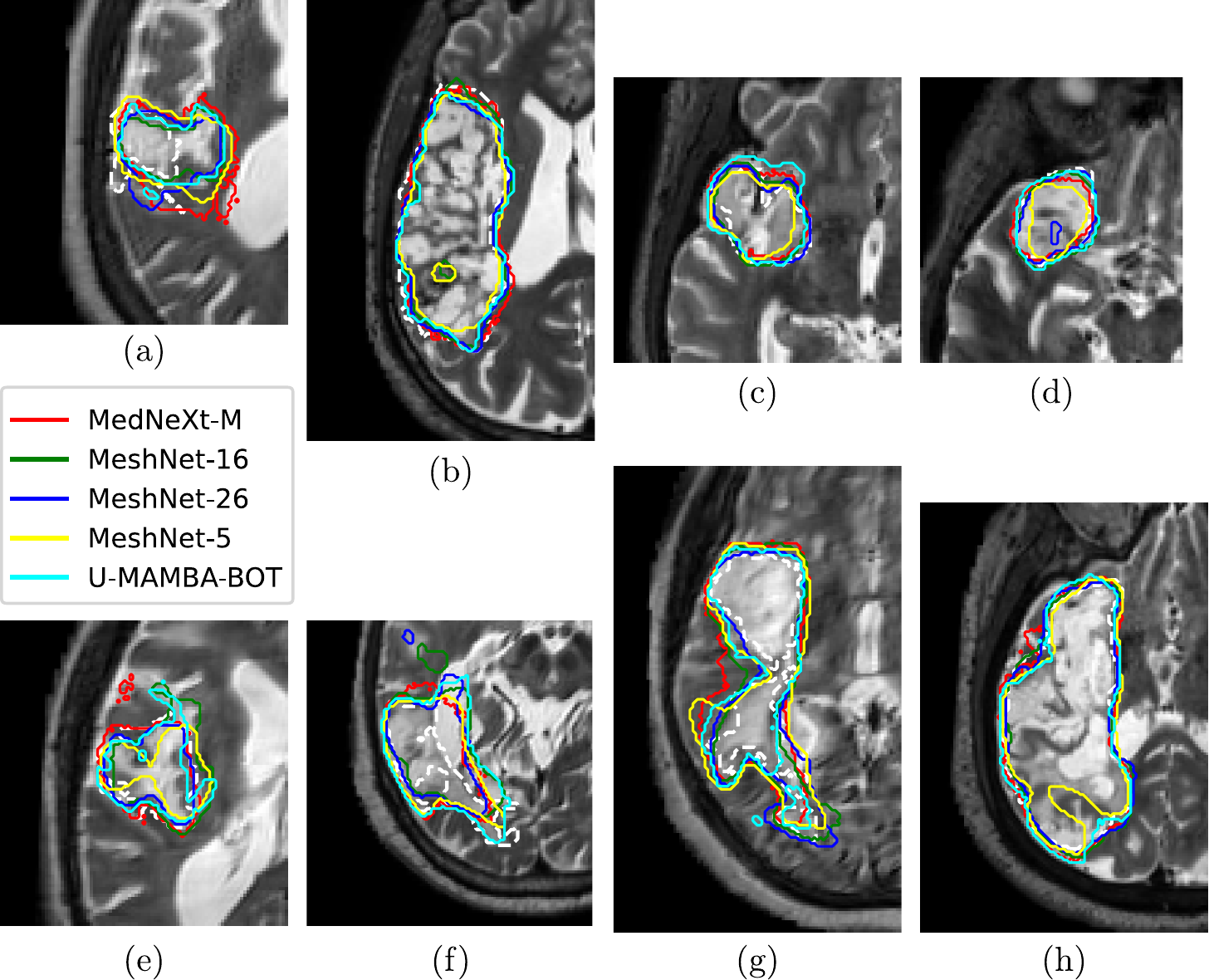}
        \vspace{-10pt}
    \caption{Comparison of chronic lesion segmentation on T2-weighted
      MRI of the ARC dataset, with color-coded contours indicating each model's output. While most models display similar alignment with lesion boundaries, subtle differences are observed. MeshNet-16 and -26 generally show fewer issues with over- or under-segmentation. In contrast, other models, including MeshNet-5, MedNeXt-M, and U-MAMBA-BOT, exhibit more frequent deviations in specific regions and variability in capturing finer lesion details.}
    \label{fig:examples}
    \vspace{-10pt}
  \end{figure}

\paragraph*{MeshNet delivers accurate lesion boundaries}

In Figure~\ref{fig:examples}, various models, including MeshNet variants (MeshNet-5, MeshNet-16, and MeshNet-26), U-MAMBA-BOT, and MedNeXt-M, are shown to demonstrate their ability to delineate chronic lesions in T2-weighted MRI images. Each color-coded contour represents the segmentation output of a specific model. The MeshNet models generally demonstrate closer alignment to lesion boundaries compared to the others, and able to capture the complex structures of chronic lesions.

We can also notice that each model exhibits distinct segmentation errors. MedNeXt-M,
represented in red, often over-segments lesion boundaries, as seen in
(b), (f), and (g), and captures non-lesion areas. MeshNet-16 and MeshNet-26 generally
follow boundaries closely but occasionally under-segment in highly
irregular regions, such as in (a) and (e), where some lesion
extensions are missed. MeshNet-5, with its minimal parameter count,
shows over- and under-segmentation in some harder areas, hence struggles with boundary precision in (e), (g) and (h). U-MAMBA-BOT, in cyan, displays inconsistent alignment and occasionally deviates from actual lesion boundaries, as seen in (f) and (g). These observations highlight the MeshNet models' relative strengths and reveal specific challenges faced by each model in delineation of chronic lesions.

\section{Conclusions}
In this work, we demonstrate that the MeshNet architecture achieves competitive performance on stroke lesion segmentation, using only 1/1000th of the parameters required by current state-of-the-art models. By implementing a multi-scale dilation pattern that mimics an encoder-decoder structure, we modified MeshNet to capture broad contextual information and fine-grained details without relying on traditional downsampling, upsampling, feature concatenation, or skip connections. Our results highlight MeshNet's strong balance of efficiency and performance in the stroke lesion segmentation task, underscoring its potential for scalable, efficient deployment in clinical and research applications such as \href{brainchop.org}{brainchop.org}~\cite{Plis2024Brainchop}. Future work will expand testing to additional datasets.

\section{Compliance with ethical standards}

This research study was conducted retrospectively using human subject data made available in open access via OpenNeuro by authors of ARC dataset~\cite{gibson2024aphasia}. Ethical approval was not required, as confirmed by the license attached with the open-access data.

\section{Acknowledgments}
This work was supported by the Nell Hodgson Woodruff School of Nursing
at Emory University (A.F., Y.B.), NIH awards
RF1NS139325 (X.H.), 2R01EB006841 (S.P.), P50-DC014664 and RF1-MH133701
(C.R.), and NSF award 2112455 (S.P.).

\bibliographystyle{IEEEbib}
\bibliography{refs}

\begin{thebibliography}{10}

\bibitem{de2015fast}
Bianca De~Haan, Philipp Clas, Hendrik Juenger, Marko Wilke, and Hans-Otto Karnath,
\newblock ``Fast semi-automated lesion demarcation in stroke,''
\newblock {\em NeuroImage: Clinical}, vol. 9, pp. 69--74, 2015.

\bibitem{roy2023mednext}
Saikat Roy, Gregor Koehler, Constantin Ulrich, Michael Baumgartner, Jens Petersen, Fabian Isensee, Paul~F Jaeger, and Klaus~H Maier-Hein,
\newblock ``Mednext: transformer-driven scaling of convnets for medical image segmentation,''
\newblock in {\em International Conference on Medical Image Computing and Computer-Assisted Intervention}. Springer, 2023, pp. 405--415.

\bibitem{ma2024u}
Jun Ma, Feifei Li, and Bo~Wang,
\newblock ``U-mamba: Enhancing long-range dependency for biomedical image segmentation,''
\newblock {\em arXiv preprint arXiv:2401.04722}, 2024.

\bibitem{ronneberger2015u}
Olaf Ronneberger, Philipp Fischer, and Thomas Brox,
\newblock ``U-net: Convolutional networks for biomedical image segmentation,''
\newblock in {\em Medical image computing and computer-assisted intervention--MICCAI 2015: 18th international conference, Munich, Germany, October 5-9, 2015, proceedings, part III 18}. Springer, 2015, pp. 234--241.

\bibitem{milletari2016v}
Fausto Milletari, Nassir Navab, and Seyed-Ahmad Ahmadi,
\newblock ``V-net: Fully convolutional neural networks for volumetric medical image segmentation,''
\newblock in {\em 2016 fourth international conference on 3D vision (3DV)}. Ieee, 2016, pp. 565--571.

\bibitem{myronenko20193d}
Andriy Myronenko,
\newblock ``{3D MRI} brain tumor segmentation using autoencoder regularization,''
\newblock in {\em BrainLes 2018}. Springer, 2019, vol. 11384 of {\em LNCS}, pp. 311--320,
\newblock In conjunction with MICCAI 2018.

\bibitem{hatamizadeh2021swin}
Ali Hatamizadeh, Vishwesh Nath, Yucheng Tang, Dong Yang, Holger~R Roth, and Daguang Xu,
\newblock ``Swin unetr: Swin transformers for semantic segmentation of brain tumors in mri images,''
\newblock in {\em International MICCAI brainlesion workshop}. Springer, 2021, pp. 272--284.

\bibitem{li2024u}
Chenxin Li, Xinyu Liu, Wuyang Li, Cheng Wang, Hengyu Liu, and Yixuan Yuan,
\newblock ``{U-KAN} makes strong backbone for medical image segmentation and generation,''
\newblock {\em arXiv preprint arXiv:2406.02918}, 2024.

\bibitem{fedorov2017end}
Alex Fedorov, Jeremy Johnson, Eswar Damaraju, Alexei Ozerin, Vince Calhoun, and Sergey Plis,
\newblock ``End-to-end learning of brain tissue segmentation from imperfect labeling,''
\newblock in {\em 2017 International Joint Conference on Neural Networks (IJCNN)}. IEEE, 2017, pp. 3785--3792.

\bibitem{gibson2024aphasia}
Makayla Gibson, Roger Newman-Norlund, Leonardo Bonilha, Julius Fridriksson, Gregory Hickok, Argye~E Hillis, Dirk-Bart den Ouden, and Christopher Rorden,
\newblock ``The aphasia recovery cohort, an open-source chronic stroke repository,''
\newblock {\em Scientific Data}, vol. 11, no. 1, pp. 981, 2024.

\bibitem{masoud2023}
Mohamed Masoud, Farfalla Hu, and Sergey Plis,
\newblock ``Brainchop: In-browser {MRI} volumetric segmentation and rendering,''
\newblock {\em Journal of Open Source Software}, vol. 8, no. 83, pp. 5098, 2023.

\bibitem{dale1999cortical}
Anders~M Dale, Bruce Fischl, and Martin~I Sereno,
\newblock ``Cortical surface-based analysis: I. segmentation and surface reconstruction,''
\newblock {\em Neuroimage}, vol. 9, no. 2, pp. 179--194, 1999.

\bibitem{xavier_bouthillier_2023_0_2_7}
Xavier Bouthillier, Christos Tsirigotis, François Corneau-Tremblay, Thomas Schweizer, Lin Dong, Pierre Delaunay, Fabrice Normandin, Mirko Bronzi, Dendi Suhubdy, Reyhane Askari, Michael Noukhovitch, Chao Xue, Satya Ortiz-Gagné, Olivier Breuleux, Arnaud Bergeron, Olexa Bilaniuk, Steven Bocco, Hadrien Bertrand, Guillaume Alain, Dmitriy Serdyuk, Peter Henderson, Pascal Lamblin, and Christopher Beckham,
\newblock ``{Epistimio/orion: Asynchronous Distributed Hyperparameter Optimization},'' 2023.

\bibitem{liew2022large}
Sook-Lei Liew, Bethany~P Lo, Miranda~R Donnelly, Artemis Zavaliangos-Petropulu, Jessica~N Jeong, Giuseppe Barisano, Alexandre Hutton, Julia~P Simon, Julia~M Juliano, Anisha Suri, et~al.,
\newblock ``A large, curated, open-source stroke neuroimaging dataset to improve lesion segmentation algorithms,''
\newblock {\em Scientific data}, vol. 9, no. 1, pp. 320, 2022.

\bibitem{Plis2024Brainchop}
Sergey~M. Plis, Mohamed Masoud, Farfalla Hu, Taylor Hanayik, Satrajit~S. Ghosh, Chris Drake, Roger Newman-Norlund, and Christopher Rorden,
\newblock ``Brainchop: Providing an {Edge} {Ecosystem} for {Deployment} of {Neuroimaging} {Artificial} {Intelligence} {Models},''
\newblock {\em Aperture Neuro}, vol. 4, sep 5 2024.

\end{thebibliography}

\end{document}